\documentclass[prl,twocolumn,superscriptaddress,showpacs,preprintnumbers]{revtex4}% Physical Review Letters
\usepackage{graphicx,epsfig}
\begin{document}
\preprint{YITP-02-55}
%%%%%%%%%%%%%%%%%%%%%%%%%%%%%%%%%%%%%%%%%%%%%%%%%%
%%%%definitions used by JF%%%%%%%%%%%%%%%%%%%%%%%
%%%%%%%%%%%%%%%%%%%%%%%%%%%%%%%%%%%%%%%%%%%%%

\def\beqra{\begin{eqnarray}} \def\eeqra{\end{eqnarray}}
\def\beqast{\begin{eqnarray*}} \def\eeqast{\end{eqnarray*}}
\def\beq{\begin{equation}}      \def\eeq{\end{equation}}
\def\be{\begin{enumerate}}   \def\ee{\end{enumerate}}

\def\gam{\gamma}
\def\Gam{\Gamma}
\def\la{\lambda}
\def\eps{\epsilon}
\def\La{\Lambda}
\def\si{\sigma}
\def\Si{\Sigma}
\def\al{\alpha}
\def\Tha{\Theta}
\def\tha{\theta}
\def\vphi{\varphi}
\def\del{\delta}
\def\Del{\Delta}
\def\ab{\alpha\beta}
\def\om{\omega}
\def\Om{\Omega}
\def\mn{\mu\nu}
\def\mun{^{\mu}{}_{\nu}}
\def\kap{\kappa}
\def\rsi{\rho\sigma}
\def\beal{\beta\alpha}
\def\til{\tilde}
\def\rta{\rightarrow}
\def\eqv{\equiv}
\def\nab{\nabla}
\def\pa{\partial}
\def\sit{\tilde\sigma}
\def\ul{\underline}
\def\indt{\parindent2.5em}
\def\nd{\noindent}
\def\rsi{\rho\sigma}
\def\beal{\beta\alpha}
% calligraphic
\def\caa{{\cal A}}
\def\cb{{\cal B}}
\def\cac{{\cal C}}
\def\cd{{\cal D}}
\def\ce{{\cal E}}
\def\cf{{\cal F}}
\def\cg{{\cal G}}
\def\cah{{\cal H}}
\def\ci{{\cal I}}
\def\cj{{\cal{J}}}
\def\ck{{\cal K}}
\def\cl{{\cal L}}
\def\cm{{\cal M}}
\def\cn{{\cal N}}
\def\cO{{\cal O}}
\def\cp{{\cal P}}
\def\car{{\cal R}}
\def\cs{{\cal S}}
\def\ct{{\cal{T}}}
\def\cu{{\cal{U}}}
\def\cv{{\cal{V}}}
\def\cw{{\cal{W}}}
\def\cx{{\cal{X}}}
\def\cy{{\cal{Y}}}
\def\cz{{\cal{Z}}}
\def\asymptotic{{_{\stackrel{\displaystyle\longrightarrow}
{x\rightarrow\pm\infty}}\,\, }} %x goes to plus minus infinity, display sty.
\def\asymptext{\raisebox{.6ex}{${_{\stackrel{\displaystyle\longrightarrow}
{x\rightarrow\pm\infty}}\,\, }$}} %x goes to plus minus infinity(text)
\def\asymptoticp{{_{\stackrel{\displaystyle\longrightarrow}
{x\rightarrow +\infty}}\,\, }} %x goes to plus infinity, display sty.
\def\asymptoticm{{_{\stackrel{\displaystyle\longrightarrow}
{x\rightarrow -\infty}}\,\, }} %x goes to minus infinity, display sty.

% nots
\def\raisenot{\raise .5mm\hbox{/}}
\def\nota{\ \hbox{{$a$}\kern-.49em\hbox{/}}}
\def\notA{\hbox{{$A$}\kern-.54em\hbox{\raisenot}}}
\def\notb{\ \hbox{{$b$}\kern-.47em\hbox{/}}}
\def\notB{\ \hbox{{$B$}\kern-.60em\hbox{\raisenot}}}
\def\notc{\ \hbox{{$c$}\kern-.45em\hbox{/}}}
\def\notd{\ \hbox{{$d$}\kern-.53em\hbox{/}}}
\def\notbd{\ \hbox{{$D$}\kern-.61em\hbox{\raisenot}}} %big D
\def\note{\ \hbox{{$e$}\kern-.47em\hbox{/}}}
\def\notk{\ \hbox{{$k$}\kern-.51em\hbox{/}}}
\def\notp{\ \hbox{{$p$}\kern-.43em\hbox{/}}}
\def\notq{\ \hbox{{$q$}\kern-.47em\hbox{/}}}
\def\notW{\ \hbox{{$W$}\kern-.75em\hbox{\raisenot}}}
\def\notz{\ \hbox{{$Z$}\kern-.61em\hbox{\raisenot}}}
\def\notpa{\hbox{{$\partial$}\kern-.54em\hbox{\raisenot}}}

\def\fo{\hbox{{1}\kern-.25em\hbox{l}}}  %raised one
\def\rf#1{$^{#1}$}
\def\bx{\Box}
\def\tr{{\rm Tr}}
\def\rmtr{{\rm tr}}
\def\dgg{\dagger}
\def\lag{\langle}
\def\rag{\rangle}
\def\bmid{\big|}
\def\vlap{\overrightarrow{\La p}} %overrightarrow 
\def\lrta{\longrightarrow} \def\lrar{\raisebox{.8ex}{$\longrightarrow$}}
\def\ON{{\cal O}(N)}
\def\UN{{\cal U}(N)}
\def\bdPh{\mbox{\boldmath{$\dot{\!\Phi}$}}}
\def\bPh{\mbox{\boldmath{$\Phi$}}}
\def\bPhs{\bPh^2}
\def\sef{S_{eff}[\sigma,\pi]}
\def\sigx{\sigma(x)}
\def\pix{\pi(x)}
\def\bph{\mbox{\boldmath{$\phi$}}}
\def\bphs{\bph^2}
\def\ex{\BM{x}}
\def\exs{\ex^2}
\def\xdot{\dot{\!\ex}}
\def\y{\BM{y}}
\def\ys{\y^2}
\def\ydot{\dot{\!\y}}
\def\pat{\pa_t}
\def\pax{\pa_x}

%\renewcommand{\thesection}{\arabic{section}}
%\renewcommand{\theequation}{\thesection.\arabic{equation}}

%%%%%%%%%%%%%%%%%%%%%%%%%%%%%%%%%%%%%%%%%%%%%%%%%
%%%%end of definitions used by JF %%%%%%%%%%%%%%%
%%%%%%%%%%%%%%%%%%%%%%%%%%%%%%%%%%%%%%%%%%%%%%%%%%
\title{Marginally Stable Topologically Non-Trivial Solitons in the 
Gross-Neveu Model}

\author{Joshua Feinberg}
\affiliation{Department of Physics, University of Haifa at Oranim,
Tivon 36006, Israel.}
\affiliation{Department of Physics, Technion, Haifa 32000, Israel.}
\affiliation{Yukawa Institute for Theoretical Physics, Kyoto
University, Kyoto 606-8502, Japan.}

\date{13 September 2002}

\begin{abstract}
We show that a kink and a topologically trivial soliton in the
Gross-Neveu model form, in the large-$N$ limit, a marginally stable
static configuration, which is bound at threshold. The energy of the
resulting composite system does not depend on the separation of its
solitonic constituents, which serves as a modulus governing the
profile of the compound soliton. Thus, in the large-$N$ limit, a kink
and a non-topological soliton exert no force on each other. 
\end{abstract}


\maketitle
The problem of finding the particle spectrum (e.g., bound
states) of quantum field theory is a major objective of
nonperturbative studies thereof. This issue may be addressed quantitatively 
in model field theories in $1+1$ space-time dimensions such as the 
Gross-Neveu (GN) model \cite{gn}, in the large $N$ limit. The GN model
(and other similar models) are particularly appealing, since they
exhibit, among other things, asymptotic freedom and dynamical mass 
generation, like more realistic four dimensional models.

One version of writing the action of the $1+1$ dimensional GN model is 
\beq\label{gnaction}
S=\int d^2x\,\left\{\sum_{a=1}^N\, \bar\psi_a\,\left(i\notpa-\si
\right)\,\psi_a 
-{1\over 2g^2}\,\si^2\right\}\,,
\eeq
where the $\psi_a\,(a=1,\ldots,N)$ are $N$ flavors of massless Dirac 
fermions, with Yukawa coupling to the scalar auxiliary field 
$\si(x)$. This action is evidently symmetric under 
the simultaneous transformations $\si\rightarrow -\si$ and 
$\psi\rightarrow\gam_5\psi$, which generate the so-called 
discrete (or ${\bf Z}\!\!\!{\bf Z}_2$) chiral symmetry of the GN model. 
The GN action has also flavor symmetry $O(2N)$, which can be seen by breaking
the $N$ Dirac spinors into $2N$ Majorana spinors. Related to this is the 
fact that the model is also invariant under charge-conjugation
\cite{dhn}. Thus, focusing on bound states of the Dirac equation 
\beq\label{diraceq}
\left[i\notpa-\si (x)\right]\,\psi = 0
\eeq
associated with the GN action, if $\psi_b(x,t) = e^{-i\om_b t}u_b(x)$ 
(with $0\leq\om_b^2< m^2$) is a bound state solution of
(\ref{diraceq}), so is its charge conjugate spinor $\psi^c_b(x,t) = 
e^{+i\om_b t}u^c_b(x)$, so that bound state 
frequencies come in pairs: $\pm\om_b$. If, however, (\ref{diraceq})
has a bound state at $\om_b=0$, it is of course unpaired, i.e.,
self-charge-conjugate. 

Performing functional integration over the grassmannian variables in 
the GN action leads to the partition function $
\cz=\int\,\cd\si\,\exp \{iS_{eff}[\si]\}$
where the bare effective action is
\beq
S_{eff}[\si] =-{1\over 2g^2}\int\, d^2x 
\,\si^2-iN\, 
\tr\log\left(i\notpa-\si\right)
\label{effective}
\eeq
and the trace is taken over both functional and Dirac indices.

The theory (\ref{effective}) has been studied in the limit 
$N\rightarrow\infty$ with $Ng^2$ held fixed\cite{gn}. In this limit 
the partition function $\cz$ is governed by saddle points of (\ref{effective}) 
and the small fluctuations around them. (In this Letter, as in
\cite{dhn}, we will consider only the leading term in the $1/N$
expansion, and thus will not compute the effect of the fluctuations
around the saddle points.) The most general saddle point
condition reads
\beq
{\delta S_{eff}\over \delta \si\left(x,t\right)}  =
-{\si\left(x,t\right)\over g^2} + iN ~{\rm tr} \left[\langle x,t | 
{1\over i\notpa
-\si} | x,t \rangle \right]= 0\,.
\label{saddle}
\eeq
In particular, the non-perturbative vacuum of the GN model is 
governed by the simplest saddle point of the path integral 
associated with it, where the composite scalar operator 
$\bar\psi\psi$ develops a space-time independent expectation value,
signaling the dynamical breakdown of the discrete chiral symmetry
by the non-perturbative vacuum. Thus, the fermions acquire mass $m$ 
dynamically.

Associated with this breakdown of the discrete symmetry is 
a topological soliton, the so called Callan-Coleman-Gross-Zee (CCGZ) 
kink \cite{ccgz,dhn,josh1}, $\sigx = m\,{\rm tanh}(mx)$, with mass
$M_{kink} = {Nm\over\pi}$ ($m$ is the dynamically generated fermion
mass). It is topology which insures the stability
of these kinks: they are the lightest topologically non-trivial
solitons in the GN model. The Dirac equation (\ref{diraceq}) in the
kink background has a single self-charge-conjugate bound state at
$\om_b =0$, which can populate at most $N$ valence fermions. Thus, it gives
rise to a multiplet of $2^N$ degenerate states, with mass equal to $M_{kink}$, 
which can be identified as the (reducible) spinor representation of $O(2N)$ 
\cite{witten,jr}. The expectation value of the fermion number operator
$N_F$ in a state in which the kink traps $n$ valence fermions is 
$n - \frac{N}{2}$, where the subtracted piece is the so-called
``fractional part'' of the fermion number \cite{witten,jr}, due to 
fluctuations of the fermion field in the topologically nontrivial
background. Thus, in the kink multiplet $-{N\over 2}\leq N_F \leq {N\over 2}$.

The GN model bears also non-topological solitons, which were
discovered by Dashen, Hasslacher and Neveu (DHN) \cite{dhn} (after the
work of \cite{ccgz}). Henceforth, we shall refer to them as ``DHN
solitons''. These non-topological solitons are stabilized 
dynamically, by trapping fermions and releasing binding energy. 
In \cite{dhn}, DHN used inverse scattering analysis \cite{faddeev} to find
static soliton solutions to the large-$N$ saddle point equations of
the GN model. (DHN also found in \cite{dhn} oscillatory, time dependent 
solutions of the saddle point equations, which we will not disucss in
this Letter.) The remarkable discovery DHN made was that {\bf all} the 
physically admissible {\em static}, space-dependent solutions of 
(\ref{saddle}), i.e., the static bag configurations in the GN model (the 
CCGZ kink being a non-trivial example of which) were {\em reflectionless}. 
That is, the static $\sigx$'s
that solve the saddle point equations of the GN model (subjected to
the obvious boundary condition $\si (\pm\infty) = \pm m$) are such  
that the reflection coefficient of the Dirac equation (\ref{diraceq})
associated with the GN action, vanishes identically.

The Dirac equation (\ref{diraceq}) in a DHN soliton 
background has a pair of charge conjugate bound states at $\pm\om_b$.  
The $O(2N)$ flavor symmetry mixes particles and antiparticles. At the
level of the Dirac equation (\ref{diraceq}) this means that we have to
consider the pair $\pm\om_b$ of bound state eigenfrequencies together
in the following way: Due to Pauli's principle, we can populate each
of the bound states $\pm\om_b$ with up to $N$ (non-interacting)
fermions. Then all the multi-particle states in which the negative 
frequency state is populated by $N-h$ fermions (i.e., has $h$ holes)
and the positive frequency state contains $p$ fermions, with $h+p =n$
fixed are degenerate in energy, and thus form a 
$C^n_{2N} = (2N)!/n!(2N-n)!$ dimensional irreducible $O(2N)$ multiplet, 
namely, an antisymmetric tensor of rank $n$ \cite{dhn}. Superficially, 
$0\leq n\leq 2N$, in accordance with Pauli's principle. However, for 
dynamical reasons, as explained in Section 3 of \cite{periodic}, only 
solitons with $0 < n < N$ are realized. The expectation value of the fermion 
number operator $N_F$ in a state in
which the DHN soliton traps $p$ particles and $h$ holes is simply $N_F = p -
h = 2p - n$, i.e., purely the naive valence contribution. (There is no 
fracional contribution due to the trivial topology.) Thus, in the 
DHN soliton multiplet $-n\leq N_F \leq n$. DHN found
that in this case $\om_b = m\,\cos\,\left({\pi n\over 2 N}\right)$.
The mass of such a soliton is $M_n = {2Nm\over\pi}\,
\sin \left({\pi n\over 2 N}\right)$ and its profile is $\sigx = \si(\infty) + 
\kappa\, {\rm tanh}\left[\kappa (x-x_0)\right] - \kappa\, {\rm tanh}\left[
\kappa \left(x-x_0\right) + {1\over 2} \log \left({m+\kappa\over m-\kappa}
\right)\right]\,,$ where $\si(\infty) = \si (-\infty) = \pm m$, 
$\kappa = m\,\sin \left({\pi n\over 2 N}\right) =\sqrt{m^2-\om_b^2}$, and 
$x_0$ is a translational collective mode. Note that both $M_n$ and $M_{kink}$ 
are of order $N\sim\frac{1}{g^2}$, as typical of soliton masses in weakly 
interacting QFT. The binding energy $B_n = nm
- M_n$ of the DHN soliton (as well as the binding energy per-fermion, 
$B_n/n$) increase with the number $n$ of trapped valence fermions
(this is the so-called ``mattress effect'' in the physics of fermion
bags). Thus, a DHN soliton is stable against decaying into a bunch of
non-interacting fundamental fermions. It is also stable against
decaying into lighter DHN bags, because any such presumed process can be 
shown to violate either energy or fermion number conservation.  
Thus, DHN solitons are stable. Note that $M_{n=N} = {2Nm\over\pi} = 
2M_{kink}$, and more over, that as $n\rightarrow N$ (i.e., 
$\kappa\rightarrow m$ and $\om_b\rightarrow 0$), the profile $\sigx$ tends 
to $\si(\infty) + m\, {\rm tanh}\,(mx) - m\, {\rm
tanh}\,[m(x+R)]\,,~R\rightarrow\infty\,.$ 
Thus, the configuration at $n=N$ is that of infinitely separated CCGZ
kink and anti-kink bound at threshold. This singular behavior occurs 
because a DHN soliton, being a topologically trivial configuration, cannot 
support a normalizable zero mode, as explained below.

The soliton solutions (both topological and
non-topological) in the GN model serve as concrete calculable examples
of fermion-bag \cite{sphericalbag,shellbag} formation. Furthermore, these 
``multi-quark''bound states of the GN model are analogous to baryons in 
QCD in the limit of large number of colors \cite{wittenlargeN}.
Since the work of DHN, these fermion bags were discussed in the
literature several other times, using alternative methods
\cite{others}. For a recent review on these and related 
matters, see \cite{thies}.

In this Letter, we show that the spectrum of the GN model contains, in
the large-$N$ limit, a composite, marginally stable topological
soliton, which may be interpreted as a kink and a non-topological
soliton bound at threshold. Furthermore, the energy of this system
does not depend on the intersoliton distance, which thus serves as a
modulus controlling the shape of the corresponding static solution of
the large-$N$ saddle point equation. 

It is a general feature of the Dirac equation (\ref{diraceq})
in the background of a static topologically non-trivial $\sigx$ 
configurations, that the spectrum contains an unpaired bound state at 
$\om=0$. For example, for kink boundary conditions ($\si (\infty) = 
-\si(-\infty) = m$), such a normalizable zero-mode is given by
$u_0\exp -\int^x \si (y)\,dy$, where $i\gamma^1 u_0 = -u_0$. 

To make our point, we have to find a topologically non-trivial static 
solution of (\ref{saddle}), which in a certain limit appears as well
separated kink and a DHN soliton. Thus, we must find a {\em reflectionless} 
$\sigx$ configuration, such that (\ref{diraceq}) has a bound state at 
$\om=0$, as required by topology, and a pair of charge conjugate bound 
states at some $\pm\om_b\neq 0$, with $\om_b$ considered a free
parameter. To this end, we apply the machinery of \cite{faddeev} 
(see also \cite{rosner}) and find the most general such reflectionless
background as 
\begin{widetext}
\beqra\label{sigmaqone2bsalt}
\sigx &=&  m + {2\kappa\over 1 + {m+\kappa\over m-\kappa}
e^{2\kappa (x-y_0)}}
\nonumber\\{}\nonumber\\
&-& 2(m+\kappa) {1+ {m+\kappa\over 
(m-\kappa)^2}\,\kappa\, e^{2m(x-x_0)} + 
{m+\kappa\over (m-\kappa)^2}\,m\,
e^{2\kappa (x-y_0)}
\over 
1+ \left({m+\kappa\over 
m-\kappa}\right)^2 e^{2m(x-x_0)} + 
\left({m+\kappa\over m-\kappa}\right)^2 e^{2\kappa (x-y_0)} + 
\left({m+\kappa\over m-\kappa}\right)^2 
e^{2m (x-x_0)+2\kappa(x-y_0)}} = \nonumber\\{}\nonumber\\
&-&\kappa\tanh\left[\kappa(x-y_0 + R)\right] \nonumber\\{}\nonumber\\
&+& \om_b{\sinh\left[m(x-x_0) +\kappa(x-y_0)+ 2\kappa R\right] + 
\sinh\left[m(x-x_0) -\kappa(x-y_0)\right]\over
e^{-\kappa R}\cosh\left[m(x-x_0) +\kappa(x-y_0)+ 2\kappa R\right] +
e^{\kappa R}\cosh\left[m(x-x_0) -\kappa(x-y_0)\right]}
\eeqra
\end{widetext}
where $\kappa = \sqrt{m^2 -\om_b^2}$, $\kappa R = \frac{1}{2}
\log\frac{m+\kappa}{m-\kappa}$, and where $x_0$ and $y_0$ are
arbitrary real parameters. The latter two quantities arise in the
inverse scattering formalism as arbitrary parameters, {\em independent} of
the parameter $\om_b$, which determine the coefficients in front of the 
asymptotic exponentially decaying {\em normalized} bound state wave
functions. Thus, the normalized bound state at $\om=0$ behaves
asymptotically as $\sqrt{2m}\,\exp -m(x-x_0)$, and the normalized bound
states at $\pm\om_b$ behave asymptotically as $\sqrt{2\kappa}\,\exp
-\kappa (x-y_0)$, as $x\rightarrow\infty$. Here, of course, $x_0$ and
$y_0$ are translational collective coordinates of the soliton $\sigx$. 
Note that $\sigx$ in (\ref{sigmaqone2bsalt}) satisfies kink boundary 
conditions: $\si (\infty) = m$ and $\si(-\infty) = m + 2\kappa
-2(\kappa+m) = -m$. Similarly, $-\sigx$ is the desired extremal
configuration with boundary conditions of an anti-kink.

The effective action $S_{eff}$ in (\ref{effective}), evaluated at the
background (\ref{sigmaqone2bsalt}), is an ordinary function 
$S_{eff}(\om_b, y_0 - x_0)$ of the parameters which determine the 
shape of (\ref{sigmaqone2bsalt}), where we have invoked 
translational invariance of (\ref{effective}). (With no loss of 
generality, we can always set one of these collective coordinates, say
$x_0$, to zero.) Minus the value of $S_{eff}(\om_b, y_0 - x_0)$ per
unit time is the rest energy, or mass $M(\om_b, y_0 - x_0)$, of the
static inhomogeneous condensate (\ref{sigmaqone2bsalt}). We still 
have to extremize $M(\om_b, y_0 - x_0)$ with respect to $\om_b$. 
As in the case of the DHN soliton, the extremal value is determined by
the number $n$ of valence particles and antiparticles which are trapped in the
bound states $\pm\om_b$. Following the same technique used by DHN 
in \cite{dhn} to calculate the mass of the DHN soliton, we find that the  
extremal value is again $\om_b = m\,\cos\,\left({\pi n\over 2
N}\right)$, and that the mass of (\ref{sigmaqone2bsalt}), which we
will refer to as the ``heavier topological soliton'' (HTS), is 
\beq\label{heavykinkmass}
M_{HTS, n} = {Nm\over\pi} + {2Nm\over\pi}\,\sin 
\left({\pi n\over 2 N}\right)\,.
\eeq
Thus, $M_{HTS,n}$ coincides with the sum of masses of a CCGZ kink and
a DHN soliton trapping $n$ valence fermions. Clearly, the $O(2N)$
quantum numbers of the HTS are those of the direct product of the
$2^N$ dimensional spinorial representation and the antisymmetric
tensor representation of rank $n$. More details of the construction of
(\ref{sigmaqone2bsalt}) and the associated extremum condition on
$\om_b$ will be given elsewhere \cite{periodic}.

Note that $M_{HTS,n}$ is {\em independent} of the remaining 
collective coordinate $y_0$. By varying $y_0$  (while keeping $\om_b$
fixed at its extremal value), we can modify the shape of $\sigx$ in 
(\ref{sigmaqone2bsalt}) without affecting the mass of the soliton. 
The translational collective coordinate $y_0$ is thus a flat direction of
the energy functional, or a modulus. In the following we will show
that the modulus $y_0$ is essentially the separation between 
a CCGZ kink and a DHN soliton which we interpret as the loosely bound
constituents of the HTS. Thus, the fact that
$\partial M_{HTS,n}/\partial y_0 =0$, means that these solitons exert 
no force on each other, whatever their separation is. 
 
This is a somewhat surprising result, since one would normally expect 
soliton-soliton interactions to be of the order $\frac{1}{g^2}\sim N$ 
in a weakly interacting field theory, which is 
consistent, of course, with what one should expect from general $\frac{1}{N}$ 
counting rules. Indeed, drawing further the analogy between the solitons 
discussed in this Letter and baryons in QCD with large $N_{color}$, the HTS 
would correspond to a dibaryon. From the general $\frac{1}{N}$ counting 
rules \cite{wittenlargeN}, the baryon-baryon interaction is expected to be of 
order $N$. Yet, due to dynamical reasons which ellude us at this point, 
the solitonic constituents of the HTS avoid these general considerations 
and do not exert force on each other.

It is straightforward to obtain the asymptotic behavior of 
(\ref{sigmaqone2bsalt}) for large $|y_0|$ (and $x_0=0$). In the limit 
$y_0\rightarrow
-\infty$, the shape of (\ref{sigmaqone2bsalt}) is that of a CCGZ kink,
with a little ``DHN bump'', centered on its left wing at
$x_{\rm{bump}}= 
y_0  + (3/4\kappa)\,\log \left({m-\kappa\over m+\kappa}\right)\simeq y_0$, of
width $1/2\kappa$ and maximum value of $m-2\om_b$. Its shape is given 
approximately by $\si (x_{\rm{bump}} + z) \simeq  -m + 2\kappa^2/(m +
\omega_b\,\cosh 2\kappa z)$. In the other limit, $y_0\rightarrow +
\infty$, (\ref{sigmaqone2bsalt}) has the shape of a CCGZ kink, with a 
little ``DHN dip'', centered on its right wing at $x_{\rm{dip}}= y_0  + 
(1/4\kappa)\,\log \left({m-\kappa\over m+\kappa}\right)\simeq y_0$, of
width $1/2\kappa$ and minimum value of $m-2\kappa
(m+\kappa-\om_b)/(m+\kappa +\om_b)$. Its shape is given 
approximately by $\si (x_{\rm{dip}} + z) \simeq  m - 2\kappa (m+\kappa)/(m +
\kappa + \omega_b\, \exp 2\kappa z) + 2\kappa \om_b /(\om_b + (m +
\kappa)\, \exp 2\kappa z)$. For $y_0$ in the range such that $\kappa
|y_0| \sim 1$, the ``DHN disturbance'' and the kink partly overlap. 
These statements are demonstrated in the figures, for the case
$m=2\kappa =1$.

\begin{figure}[htb]
%\vspace{-2cm}
\epsfig{file=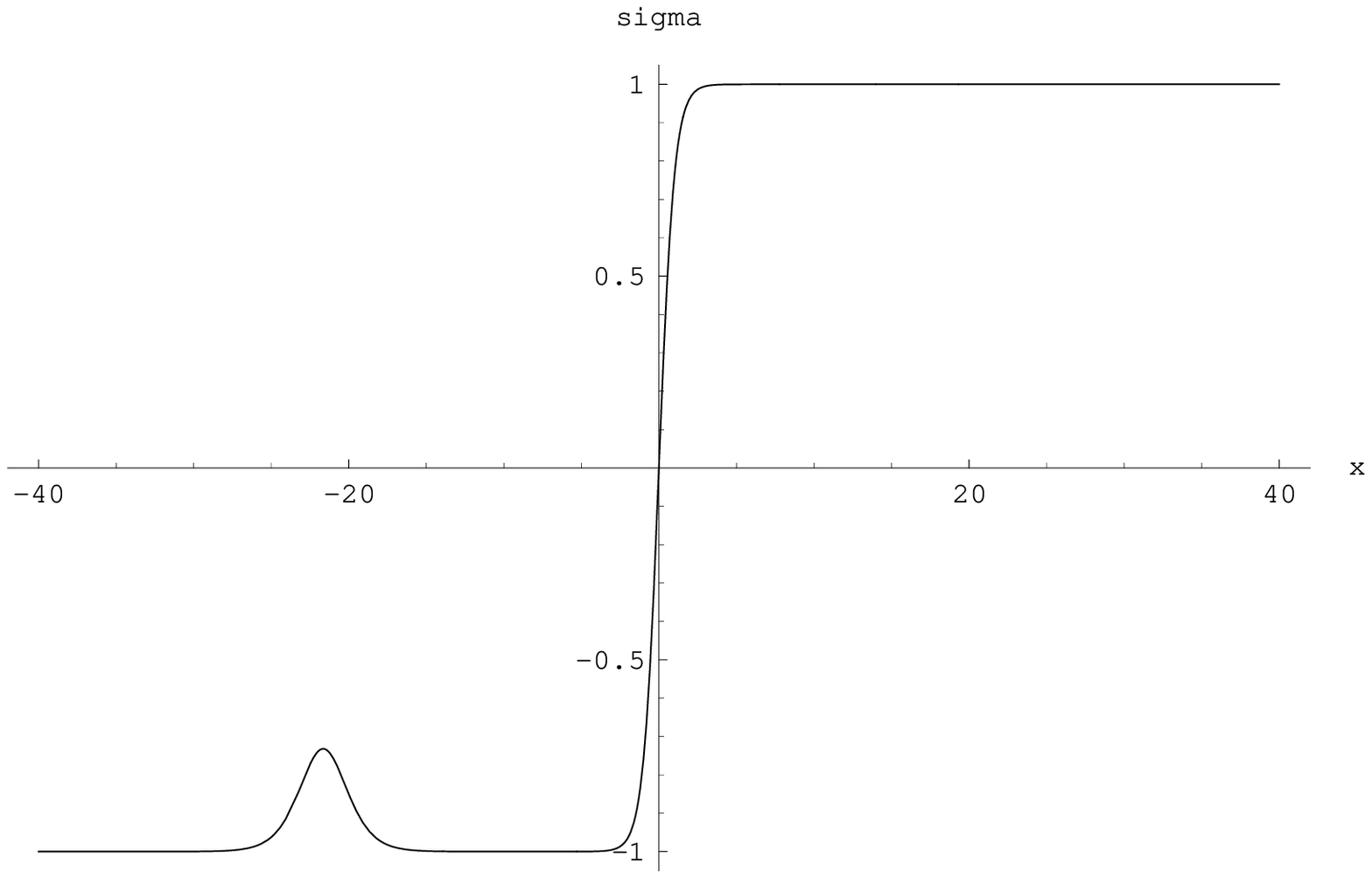,height=6.5cm,width=8cm,angle=0}
\caption{The soliton (\ref{sigmaqone2bsalt}) for $m=2\kappa = 1;
2\kappa y_0  = -20$ 
%\vspace{-0.5cm}
\label{minus20}}
\end{figure}

\begin{figure}[htb]
%\vspace{-2cm}
\epsfig{file=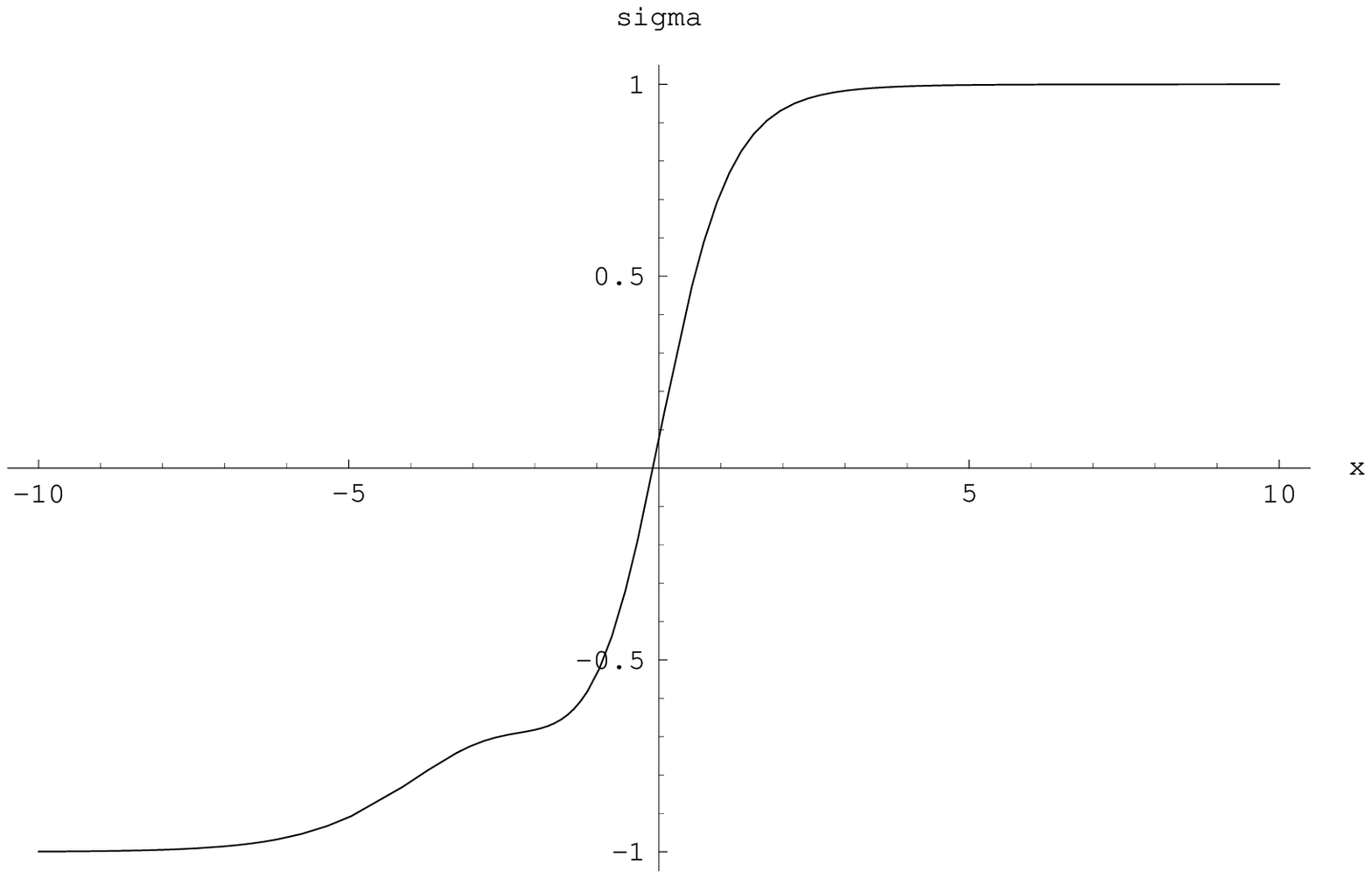,height=6.5cm,width=8cm,angle=0}
\caption{The soliton (\ref{sigmaqone2bsalt}) for $m=2\kappa = 1;
2\kappa y_0  = -1$ 
%\vspace{-0.5cm}
\label{minus1}}
\end{figure}

\begin{figure}[htb]
\epsfig{file=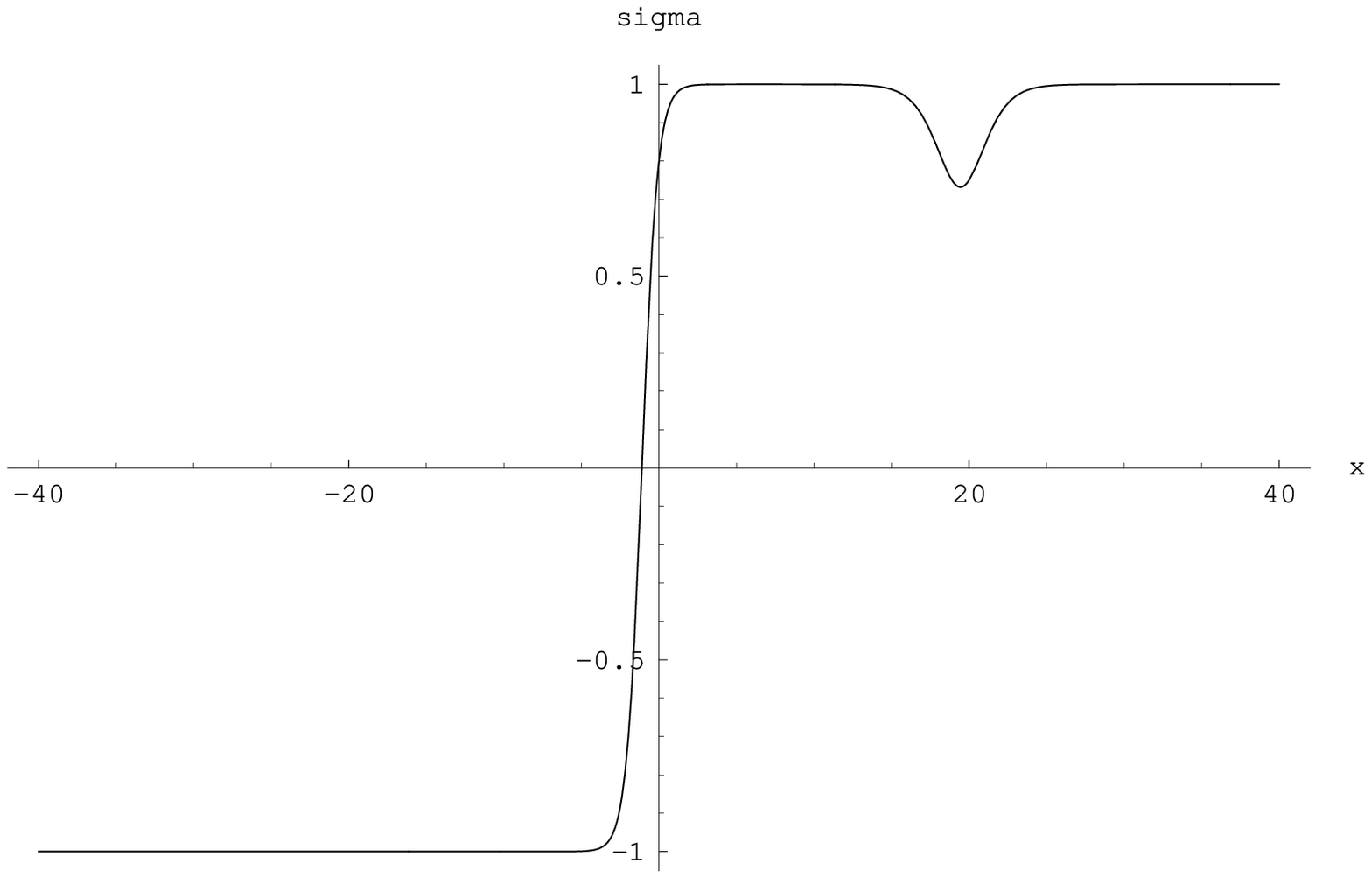,height=6.5cm,width=8cm,angle=0}
%\vspace{-2cm}
%\includegraphics[bb=18 18 552 451, width=\textwidth]{kinkplus20.eps}
\caption{The soliton (\ref{sigmaqone2bsalt}) for $m=2\kappa = 1;
2\kappa y_0  = +20$ 
%\vspace{-0.5cm}
\label{plus20}}
\end{figure}
%Finally, in the special case in which $2mx_0 = 2\kappa y_0 = \log 
%\left({m+\kappa\over m-\kappa}\right)$, $\sigx$ in (\ref{sigmaqone2bsalt})
%is a kink-like odd function. (For example, for $\kappa=m/2$ one obtains
%in this case $\sigx = m \tanh\,(mx/2)$, i.e., a deformation of the
%CCGZ kink which is spread out in space as twice as much as the CCGZ kink.)

The limit $n\rightarrow N$ is of some interest. Strictly speaking, there is
no $HTS$ with $n=N$, since at $n=N$ the pair of bound states at $\pm\om_b$
of the Dirac equation (\ref{diraceq}) would coincide with the bound state
at $\om=0$, which already exists due to non-trivial topology. Clearly, 
such a degeneracy cannot occur in the spectrum in one spatial dimension. 
(For more details see Sections 3.1.2 and B.3.2 in \cite{periodic}.) 
However, it is possible to study HTS's with $n$ arbitrarily close to $N$. 
In this case, $\kappa\rightarrow m$, with $R\rightarrow \infty$ 
in (\ref{sigmaqone2bsalt}). Thus, for $|x|, |x_0|, |y_0| << R$, 
(\ref{sigmaqone2bsalt}) tends in this limit to 
\beq\label{sigmaqone2bsntoN}
\sigx = m{1-e^{-2m(x-y_0)}-e^{-2m(x-x_0)}
\over 1+e^{-2m(x-y_0)}+e^{-2m(x-x_0)}}
\eeq
(where for clarity of presentation we have reinstated the parameter $x_0$ 
into the expression for $\sigx$). In the asymptotic region $1<< m|x_0-y_0|~
 (<< mR)$ (\ref{sigmaqone2bsntoN}) simplifies further, and appears as a 
kink  $m\tanh[m(x-x_{\rm max})]$, located at $x_{\rm max} = 
{\rm max}\{x_0, y_0\}$. This clearly has mass $M_{kink} = {Nm\over\pi}$, but 
according to (\ref{heavykinkmass}), $M_{HTS,n\simeq N}$ should tend to 
${3Nm\over\pi} = 3M_{kink}$. The extra mass $2M_{kink}$ corresponds, of 
course, to the kink-anti-kink pair which receded to spatial infinity.

Finally, we must settle the important issue of stability of the HTS. 
Due to conservation of the topological charge $q = (\si(\infty) 
-\si(-\infty))/2m$, the final static configuration will obey the same
kink boundary conditions as  (\ref{sigmaqone2bsalt}). The HTS is too
light to decay into a configuration of CCGZ kink-kink-antikink
(see (\ref{heavykinkmass})). Thus, it can only decay into a CCGZ kink plus
a bunch of lighter DHN bags and/or free fermions. The binding energy
$B_n = nm + M_{kink} - M_{HTS,n }$ of the HTS coincides with that of a
DHN bag with the same quantum number $n$. Thus, similarly to the
latter, the HTS is stable against evaporation into a CCGZ kink and a
cloud of non interacting fermions. Decay of the HTS (with $O(2N)$ quantum
number $n$ ) into a CCGZ kink and a DHN bag (of quantum number $n'$)
is almost totally forbidden: Energy conservation obviously requires 
$0 < n'\leq n$. Fermion number conservation, on the other hand, requires 
(see \cite{periodic} for more details) $n\leq n' < N$, i.e.,
the complimentary set of the possible range $(0,N)$ for $n'$. The two 
conservation laws are compatible only at $n'=n$. Thus, the mass of the
decay products equals the mass of the parent HTS, and the allowed channel
in this case has no phase space. Finally, one can show, using elementary 
considerations as above, that energy conservation and fermion number 
conservation strictly forbid decay of the HTS into a lighter HTS and a DHN 
soliton, or into a CCGZ kink plus any number of DHN solitons, or into any 
final state containing the time dependent solitons discovered by DHN in 
\cite{dhn}. Since the only allowed channel has no phase space, the HTS is 
marginally stable. This must be a manifestation of the fact that the 
translational collective coordinate $y_0$ is a flat direction of the energy
functional.

\begin{acknowledgments}
I am happy to thank Y. Frishman, M. Karliner, M. Moshe and R. Sasaki 
for useful discussions, and N. Andrei for correspondence. Also, I
thank the members of the high energy theory group at the Yukawa
institute, where this work was completed, for their cordial
hospitality. This work has been supported in part by the Israeli
Science Foundation.
\end{acknowledgments}

\typeout{References}

\end{document}